\documentclass[a4paper,aps,pre,floatfix,preprint,nofootinbib]{revtex4}
\usepackage[dvips]{epsfig}
\usepackage[dvips]{graphicx}
\usepackage{latexsym,amsmath,verbatim}
\usepackage{color}

\begin{document}

\title{Modeling the emergence of universality in color naming patterns}

\vspace{3cm}

\author{Andrea Baronchelli\footnote{To whom correspondence should be addressed: andrea.baronchelli@upc.edu}}
\affiliation{Departament de Fisica i Enginyeria Nuclear,
  Universitat Politecnica de Catalunya, Campus Nord B4, 08034
  Barcelona, Spain}

\author{Tao Gong}
\affiliation{Department of Linguistics, Max Planck Institute for
  Evolutionary Anthropology, Deutscher Platz 6, 04103 Leipzig, Germany}

\author{Andrea Puglisi}
\affiliation{CNR-INFM-SMC and Dipartimento di Fisica, Sapienza Universita' di Roma,
  Piazzale Aldo Moro 5, 00185 Roma, Italy}

\author{Vittorio Loreto}
\affiliation{Dipartimento di Fisica, Sapienza Universita' di Roma,
  Piazzale Aldo Moro 5, 00185 Roma, Italy and Fondazione ISI, Torino, Italy}

\begin{abstract}
  The empirical evidence that human color categorization exhibits some
  universal patterns beyond superficial discrepancies across
  different cultures is a major breakthrough in cognitive science. As
  observed in the World Color Survey (WCS), indeed, any two groups of individuals
  develop quite different categorization patterns, but some universal
  properties can be identified by a statistical analysis over a large
  number of populations. Here, we reproduce the WCS in a numerical
  model in which different populations develop
  independently their own categorization systems by playing elementary
  language games.  We find that a simple perceptual constraint shared
  by all humans, namely the human Just Noticeable Difference (JND), is
  sufficient to trigger the emergence of universal patterns that
  unconstrained cultural interaction fails to produce.  We test the
  results of our experiment against real data by performing the
  same statistical analysis proposed to quantify the universal
  tendencies shown in the WCS [Kay P \& Regier T. (2003) Proc. Natl. Acad. Sci. USA 
  100: 9085-9089], and obtain an excellent quantitative
  agreement. This work confirms that synthetic modeling
  has nowadays reached the maturity to contribute significantly
  to the ongoing debate in cognitive science.
\end{abstract}

\maketitle

The finding that color naming patterns present some conserved features
across cultures~\cite{berlinkay} is a milestone in the debate on the
existence of universals in human
categorization~\cite{gardner1985msn}. The data collected in the World Color 
Survey (WCS)~\cite{cook2005world}, extending the pioneering work by Berlin 
and Kay~\cite{berlinkay}, provide empirical
evidence in favor of the fact that categorization is not simply a
matter of conventions, but rather depends on the physiological
and cognitive features of the categorizing subjects, in contrast with
previous theories according to which categories are
arbitrarily defined by different cultures~\cite{whorf56}.  
Over the years, the existence of universals in color categorization has
become generally accepted~\cite{gardner1985msn,lakoff-women,taylor2003lc,murphy2004bbc}, 
though it has been the subject of strong controversy, and a debate is still ongoing~\cite{saunders1997tnc,Davidoffetal1999,Robersonetal2000,Robersonetal2005,roberson2007cvc}.
% Even though the existence of universals in color categorization has gained
% ground over the years~\cite{gardner1985msn,lakoff-women,taylor2003lc,murphy2004bbc},
% the issue has been the subject of strong controversy, some of which
% are part of a still ongoing
% debate~\cite{saunders1997tnc,Davidoffetal1999,Robersonetal2000,Robersonetal2005,roberson2007cvc}.
Recently, however, a set of statistical tests have proved
quantitatively that the WCS data do in fact contain clear signatures
of universal tendencies in color naming, both across industrialized
and non-industrialized languages~\cite{kay2003rqc}.  In any case, the
WCS maintains a central role and its data, as a fundamental (and almost unique)
experimental repository, are still under constant
scrutiny, as shown by the continuous flow of publications
related to them (see, for instance,
\cite{kay2003rqc,maclaury1987cce,Regieretal2005,lindsey103ucn,Regieretal2007}).

Color categorization represents a case study in a wide debate on the
origins, meanings and properties of categorization
systems~\cite{taylor2003lc,murphy2004bbc}. In recent years,
mathematical and computational models have been designed to explore
the roles of various hypotheses concerning these issues, checking their 
implications in simplified, yet hopefully transparent, synthetic
experiments~\cite{frankfurt}. Generally speaking, a model can show
whether its assumptions are internally coherent,
whether they can produce the claimed effects, and above all, whether
they are sufficient in principle to generate a given
phenomenon. Although a model cannot prove if those assumptions are
empirically valid, it can test the available hypotheses, suggest
new perspectives, and help to ask more focused
questions~\cite{frankfurt}. Color categorization has been used as a
reference problem also in computational studies that investigate how much
language and perceptually grounded categories influence each other and
how a population of individuals establish a shared repertoire of
categories. Pioneering work in this direction has shown that purely
cultural negotiation in the form of iterated Language
Games~\cite{wittgenstein53english} can lead to a co-evolution of
categories and their linguistic labels~\cite{steels2005cpg,Belpaeme_Bleys2005} in
a population of individuals. This line of research has been
subsequently extended, and complex systems methods have demonstrated
that cultural interaction is able to yield a finite number of categories 
even when the perceptual space is continuum, as in
the case of color perception~\cite{cg_pnas}.
%by adopting methods from complex systems, based
%on which cultural interaction is shown to be able to yield a finite
%number of shared categories from a continuum perceptive space, as in
%the case of color perception~\cite{cg_pnas}. 
A different point of view has been adopted in the framework of the Iterated Learning
Model~\cite{kirby2002bae,smith2003ilf}, in which a population is
modeled as a chain of individuals, each learning from the output of
the previous generation and providing the input to the subsequent
one~\cite{kirby2007iac}. In this context, it has been suggested that universals in
categorization may originate from the presence of unevenly distributed
salient color foci in the perceptual space~\cite{dowman2007ect}. Finally, the
Evolutionary Game Theory~\cite{nowak2006ed} approach
%has contributed to our understanding of color categorization, 
has focused mainly on the roles played by different individual features (being
linguistic, psychological or physiological) on the shared color
categorization~\cite{komarova2007emc}, such as the influence of few
abnormal observers on the whole categorization
system~\cite{komarova2008pha}.

Employing an \textit{in silica} experiment, in this paper 
we show that cultural transmission can induce universal patterns in
color categorization, provided that some basic properties of human
perceptual system are considered. We generate ``synthetic'' languages
via an agent-based model~\cite{cg_pnas} that simulates a number of
independent groups of interacting individuals. We identify 
universal patterns in color naming among groups whose members are
endowed with the human Just Noticeable Difference (JND) function
describing how the resolution power of human vision varies according
to the frequency of the incident light. These results are tested
against the experiment in which individuals perceive the spectrum
homogeneously. Strikingly, following the same analysis
of~\cite{kay2003rqc}, we find that the difference between these two
types of languages is in excellent agreement with the difference
between the experimental and randomized data measured by Kay and
Regier in their work based on the WCS~\cite{kay2003rqc}. Such an agreement is remarkable, 
considering the rather minimal input introduced: except for the JND curve, our experiment is
blind to any other properties of the real world or real human beings.

\section{The Category Game Model}

The computational model used in this experiment, introduced
in~\cite{cg_pnas}, involves a population of $N$ artificial
agents. Starting from scratch and without pre-defined color
categories, the model dynamically generates, via a number of
``games'', a pattern of linguistic categories for the visible light
spectrum highly shared in the whole population. The model has the
advantage of incorporating an extremely low number of parameters,
basically the number of agents $N$ and the JND curve $d_{min}(x)$
(detailed in the Methods), compared with its rich and realistic
output.

For the sake of simplicity and not loosing the generality for the
purpose of analysis, color perception is reduced to a single
analogical continuous perceptual channel, each stimulus being a real
number in the interval $[0,1)$ that represents its normalized,
rescaled wavelength. A categorization pattern is identified as a
partition of the interval $[0,1)$ into sub-intervals, or perceptual
categories. Individuals have dynamic inventories of form-meaning
associations that link perceptual categories with their linguistic
counterparts, i.e., basic color terms, and these inventories evolve
through elementary language games~\cite{wittgenstein53english}. At
each time step, two players (a speaker and a hearer) are chosen
randomly from the population and a scene of $M\ge 2$ stimuli is
presented to them. Any two of these stimuli cannot appear at a
distance smaller than $d_{min}(x)$, where $x$ is the value of either
of the two. In this way, the JND is implemented in the model. Based on
the presented stimuli, the speaker discriminates the scene, if
necessary refines its perceptual categorization, and utters the color
term associated to one of the stimuli (the topic). The hearer tries to
guess the topic, and based on the outcome (success or failure) both
individuals rearrange their form-meaning inventories (further details of this
process are given in the Methods, and the complete algorithm of the
Category Game is provided in the Supporting Information ($SI$)). New
color terms are invented every time a new category is created for the
purpose of discrimination, and these terms can diffuse through the
population in successive games. At the beginning, all individuals have
only the perceptual category $[0,1)$ with no associated name.

During the first phase of the evolution, the pressure of
discrimination makes the number of perceptual categories increase: at
the same time, many different words are used by different agents for
some similar categories. This kind of synonymy reaches a peak and then
dries out, in a way similar to the well-known Naming
Game~\cite{steels1995,baronchelli_ng_first,baronchelli2008dan}. When
on average only one word is recognized by the whole population for
each perceptual category, the second phase of the evolution
intervenes. During this phase, words expand their reference across
adjacent perceptual categories, joining these categories to form new
``linguistic categories''. The coarsening of these categories becomes
slower and slower, with a dynamic arrest analogous to the physical
process in which supercooled liquids approach the glass
transition~\cite{mezard1987sgt}. In this long-lived, almost stable
phase, usually after $10^4$ games per player, the linguistic
categorization pattern has a $90\%$ to $100\%$ degree of sharing among
individuals and remains stable for a long plateau phase whose duration
diverges with the population size (lasting for instance $10^5 \sim
10^6$ games per player~\cite{cg_pnas} for the population sizes
considered) . We consider the pattern corresponding to this plateau
phase as the ``final pattern'' generated by the model, i.e. the most
relevant for comparison with human color categories. If one waits for
a much longer time, the number of linguistic categories will drop. 
This unrealistic effect is caused by
the extremely slow diffusion of category boundaries\footnote{At for the Category
  Game, categories can be equivalently described in terms of
  boundaries or prototypes, without any difference~\cite{cg_pnas}.},
which is due to finite-size effects, occurring at
longer and longer times as the population size increases. Nonetheless,
since the comparison with the real world is much less accessible on
such a long time-scale, we are not interested in the behavior of the
model in this phase (see also the discussion on the relevance of
pre-asymptotic states in categorization models in~\cite{frankfurt}).

The shared pattern in the long-lived, stable phase between $10^4$ and
$10^6$ games per player is the focus of the experiment described in
the following section, see Figure~1 for an example. It is remarkable,
as already observed in~\cite{cg_pnas}, that the number of linguistic
color categories achieved in this phase is of the order of $20 \pm
10$, even if the number of possible perceptual categories ranges
between $100$ and $10^4$ and the number of agents ranges between $10$
and $1000$. For this reason, the mechanism of spontaneous emergence of
linguistic categories in this model is relevant for the problem of
linguistic categorization in continuous spaces (such as the perceptual
space of colors) where no objective boundaries are present.

\section{The Numerical World Color Survey}

The aim of our study is to replicate \textit{in silica} the WCS by
performing a Numerical World Color Survey (NWCS). To this purpose, we
generate ``worlds'' made of isolated populations. Each population is
the outcome of a run of the model with $N$ individuals, and each
``world'' collects $50$ such populations (the logical scheme of this
experiment is shown in Fig.~2). The sequence of games in each run is
random, thus making each evolution history unique and each final
shared pattern of linguistic color categories different across
populations. Two classes of ``worlds'' are created:  ``human
worlds'' are obtained by endowing the individuals with the human JND
function, while  ``neutral worlds'' are obtained by using a
uniform JND, i.e., $d_{min}(x)=0.0143$, which is the average value of
the human JND (as it is projected on the $[0,1)$ interval).

In all cases, as showed in~\cite{cg_pnas}, each population develops a
shared repertoire of roughly $10-20$ linguistic categories in the
stable phase: this number of linguistic categories is weakly dependent
on $N$, and the value $N=50$ is a good compromise to obtain
representative results without increasing too much the length of
simulations. The robustness of our findings with respect to different population
sizes, as well as to changes in other parameters of the model, is well verified, as reported in the
$SI$. The hypothesis we test here is that the similarity between the
linguistic patterns developed in the ``human worlds'' is higher on
average than the one observed in the ``neutral worlds''. To this aim,
we compute, for each ``world'', the quantity $D$ defined to measure
the dispersion of patterns of color terms in the WCS~\cite{kay2003rqc}
(see the Methods for its definition). Following the same procedure
used in the WCS, we define the representative point of each linguistic
category as its central point (see $SI$ for the details of our
procedure).

The analysis of the WCS data has showed that the patterns collected in
the survey are less dispersed (i.e., more clustered) than their
randomized counterparts, thus proving the existence of universality in
color categorization. Our simulations consider the data obtained from
the ``neutral worlds'' rather than the randomized data, but the
meaning of the test is analogous and represents a standard procedure
in statistical analysis~\cite{dagostini}: when the data in a set are
believed to present some kind of correlation, the hypothesis is tested
against uncorrelated data sets.
%i.e., in order to assess
%that the data in a set present some kind of correlation, one needs to
%test these data against the data in sets that are known to be
%uncorrelated. 
In analogy to the WCS experiment, the randomness hypothesis
in the NWCS for the test-cases (the ``neutral worlds'') is supported
by symmetry arguments: in each neutral simulation there is no
breakdown of translational symmetry, which is the main bias in the
``human world'' simulations.

Our main results are presented in Figure~3. Since the dispersion $D$
defined in~\cite{kay2003rqc} depends on the
number of languages, the number of colors, and the space units used,
it is convenient to divide every measure of $D$ in the NWCS by the
average value obtained in the ``human world'' simulations, and every
measure of $D$ from the WCS experiment by the value obtained in the
original (non-randomized) WCS analysis (as in
~\cite{kay2003rqc}). Then, both the average of the ``human worlds''
and the value based on the WCS data are represented by $1$ in
Figure~3, as pointed out by the big black arrow. In the same plot, the
probability density of observing a value of $D$ in the ``neutral
world'' simulations is also shown by the red histogram bars. The
probability density $\rho(x_i)$ equals to the percentage $f(x_i)$ of
the observed measure in a given range $[x_i-\Delta/2,x_i+\Delta/2]$
centering around $x_i$, divided by the width of the bin $\Delta$,
i.e., $\rho(x_i)=f(x_i)/\Delta$. This procedure allows for a comparison
between the histogram coming from the NWCS and that obtained in the
study on the WCS ~\cite{kay2003rqc}, where the bins have a different
width. We also import (by digitalization) the data reported in the
histogram of the randomized datasets in Figure 3a
of~\cite{kay2003rqc}. To be consistent with the above procedure, 
the abscissa is normalized by the value of the
non-randomized dataset and the frequencies are rescaled by the width of
the bins.

Figure~3 illustrates two remarkable results. First, the Category Game
Model informed with the human $d_{min}(x)$ JND curve produces a class
of ``worlds'' that has a dispersion lower than and well distinct from that
of the class of ``worlds'' endowed with a non-human, uniform
$d_{min}(x)$. Second, the ratio observed in the NWCS between the
average dispersion of the ``neutral worlds'' and the average
dispersion of the ``human worlds'' is $D_{neutral} / D_{human} \sim
1.14$, very similar to the one observed between the randomized
datasets and the original experimental dataset in the WCS.

In summary, the color categories emerged in the ``human worlds'' in
the NWCS are significantly less dispersed than those emerged in the
``neutral worlds''\footnote{This remains true when taking account of
  the whole dispersion histogram of the ``human worlds'', as discussed
  in the ($SI$).}, and this difference agrees quantitatively with that
observed in the WCS, where human languages are considered. Also, in
the $SI$, we show that even changing the population size, the number
$M$ of objects in a scene, or the time (measured in terms of games per
agent) at which the categorization system is observed, our results
agree quantitatively with the ones reported in~\cite{kay2003rqc},
within an error of at most $10\%$.
Considering the huge degree of reduction and simplification that
separates the Category Game Model from the human language, 
this finding is definitely remarkable.
%he agreement we find in our \textit{in silica} experiment is absolutely
%remarkable.

\section{Discussion and conclusion}

Through an \textit{in silica} experiment, we have shown that
independent groups of interacting agents incorporating a single human,
presumably biological, constraint (the human JND function) end up
developing categorization systems that exhibit universal
properties similar to those observed in the WCS. We have also pointed out
that replacing the human JND function with the uniform JND produces a
similar effect of an \textit{a-posteriori} randomization on the WCS data,
which is illustrated by the quantitative agreement between the results
obtained in our experiment and those extracted from the WCS data
in~\cite{kay2003rqc}. Taken as a whole, our findings corroborate 
previous evidence in favor of the existence of universality in color
naming systems~\cite{kay2003rqc,lindsey2006universality}. Moreover,
they suggest that the human perceptual (i.e., visual) system can be
responsible for, or at least involved in, the emergence of such
universal properties of categorization
systems~\cite{jameson1997s,Regieretal2007}. More precisely, the NWCS
proves that purely cultural interaction among individuals sharing an
elementary perceptual bias is sufficient to trigger the emergence of
the universal tendencies observed in human categorization. Although
the bias does not affect the properties of the shared categorization
system in a deterministic way, it is responsible for subtle
similarities that can be revealed by a statistical analysis on a large
number of different populations.

Our work also demonstrates that computational approaches have nowadays
reached good maturity, since the multi-agent model presented here (i)
incorporates straightforwardly a real feature of human perceptual
system (i.e. the human hue-JND), and produces results (ii) testable
against and (iii) in quantitative agreement with the empirical
data. In addition, since the model is simply designed, there is a
particularly transparent connection between the incorporated
hypothesis and the generated results. Future work can further
enrich the present picture, for example by considering a
multidimensional perceptual channel or the impact of differences among
individuals on the emergent category system, in the spirit
of~\cite{komarova2008pha}. Moreover, the close tie between our study
and human perception, as discussed above, can help to inspire new
experiments and to design and analyze human or artificial
communication systems~\cite{cattuto2007sda,pnas_cattuto_etal_2009}. In
conclusion, we believe that the results presented here not only
contribute to the debate on the origins of universals in
categorization, but also stimulate new efforts towards the growth of a
computational cognitive science.

\section{Methods}

\subsection{The WCS and the dispersion measurement}

A first survey was run on 20 languages in $1969$ by
P. Kay and B. Berlin ~\cite{berlinkay}. From 1976 to 1980, the World
Color Survey was conducted by the same researchers along with
W. Merrifield. Since 2003, the data have been made public on the
website {\tt http://www.icsi.berkeley.edu/wcs}. These data concern the
basic color categories in $110$ languages without written forms and
spoken in small-scale, non-industrialized societies. On average, $24$
native speakers of each language were interviewed. Each informant had
to name each of $330$ color chips produced by the Munsell Color
Company that represent $40$ gradations of hue and maximal saturation,
plus $10$ neutral color chips (black-gray-white) at $10$ levels of
value. These chips were presented in a predefined, fixed random order.

Recently, Kay and Regier~\cite{kay2003rqc} performed the following
statistical analysis. After a suitable transformation, they identified
the most representative chip for each color in each language as a
point in the CIEL*a*b color space, where an Euclidean distance is
defined. In order to investigate whether these points are more
clustered across languages than would be expected by chance, they
defined a dispersion measure on this set of languages $S_0$
$$ D_{S_0}= \sum_{l,l^* \in S_0} \sum_{c \in l} \mbox{min}_{c^* \in
  l^*} \mbox{distance}(c,c^*),
$$
where $l$ and $l^*$ are two different languages, $c$ and $c^*$ are two
basic color terms respectively from these two languages, and
$\mbox{distance}(c,c^*)$ is the distance between the points in
CIEL*a*b space that represent the two colors. To give a
meaning to the measured dispersion $D_{S_0}$, Kay and Regier created
some ``new'' datasets $S_i$ ($i=1,2,..,1000$) by random rotation of
the original set $S_0$, and measured the dispersion of each new set
$D_{S_i}$. The ``human'' dispersion appears to be distinct from the
histogram of the ``random'' dispersions with a probability larger than
$99.9\%$. As shown in Figure 3a of~\cite{kay2003rqc}, the average
dispersion of the random datasets, $D_{neutral}$, is $1.14$ times
larger than the dispersion of human languages.

\subsection{The Just Noticeable Difference}

As shown in~\cite{long2006ssn}, humans view the world in a non-uniform
way; they have different perceptual precisions for stimuli with
different wavelengths from a given continuous hue
space. Psychophysiologists define the Just Noticeable Difference (JND) as
a function of wavelength to describe the minimum distance at which two
stimuli from the same scene can be discriminated. In principle, this
parameter can either be taken as constant across the whole perceptual
interval or be modulated to account for the regions with different
resolution powers. Based on~\cite{long2006ssn}, we build up a human
JND function as shown in Figure~3, compared with the uniform JND.

\subsection{Details of the simulated model}

In the Category Game~\cite{cg_pnas}, at each time step two agents (a
speaker and a hearer) are picked up to conduct a language game, during
which the mechanisms for interaction and bargaining are as follows. A
scene containing $M \ge 2$ stimuli is presented to these agents; each
pair $x,y$ of the stimuli must be at a distance larger than
$d_{min}(x)$. One of the stimuli, known to the speaker only, is the
topic. The speaker checks if there is a perceptual category in which
only the topic lies. If two (or more) stimuli lie in the category assigned 
to the topic, the speaker divides it into two (or more) new categories, each
inheriting the words associated to the original category and acquiring
a new word; this process is called
``discrimination''~\cite{steels2005cpg,cg_pnas}. After that, the
speaker utters the most relevant name of the category containing the
topic only (the most relevant name is the last name used in a winning
game or the new name if this category is newly created). If the hearer
does not have a category with that name, the game fails. If the hearer
recognizes the name and has some stimuli in a category that is
associated with it in her inventory, she picks randomly one of
these stimuli (if M is not large, the hearer typically has a single
candidate, see~\cite{cg_pnas}). If the picked candidate is the topic,
the game succeeds; otherwise, it fails. In case of failure, the
hearer learns the name used by the speaker for the topic's category;
in case of success, that name becomes the most relevant name for
the used category in both agents and all the other competing
names in that category are removed. The full algorithmic
description of the model is provided in the $SI$.

\begin{acknowledgments}
  The authors are indebted to Romualdo Pastor-Satorras for helpful discussions
  on the numerical simulations.
  A. Baronchelli acknowledges support from the Spanish Ministerio de
  Ciencia e Innovaci\'{o}n through the Juan de la Cierva Grant funded by the 
  European Social Fund and from Project FIS2007-66485-C02-01 
 (Fondo Europeo de Desarrollo Regional). T. Gong acknowledges support from the Alexander von
  Humboldt Foundation in Germany. A. Puglisi acknowledges support by
  the Italian MIUR under the FIRB-IDEAS grant RBID08Z9JE. V. Loreto
  acknowledges support from the TAGora and ATACD projects funded by
  the European Commission under the contracts IST-34721 and 043415.
\end{acknowledgments}

\clearpage

\begin{figure}
 \begin{center}
 \includegraphics[width=14cm,clip=true]{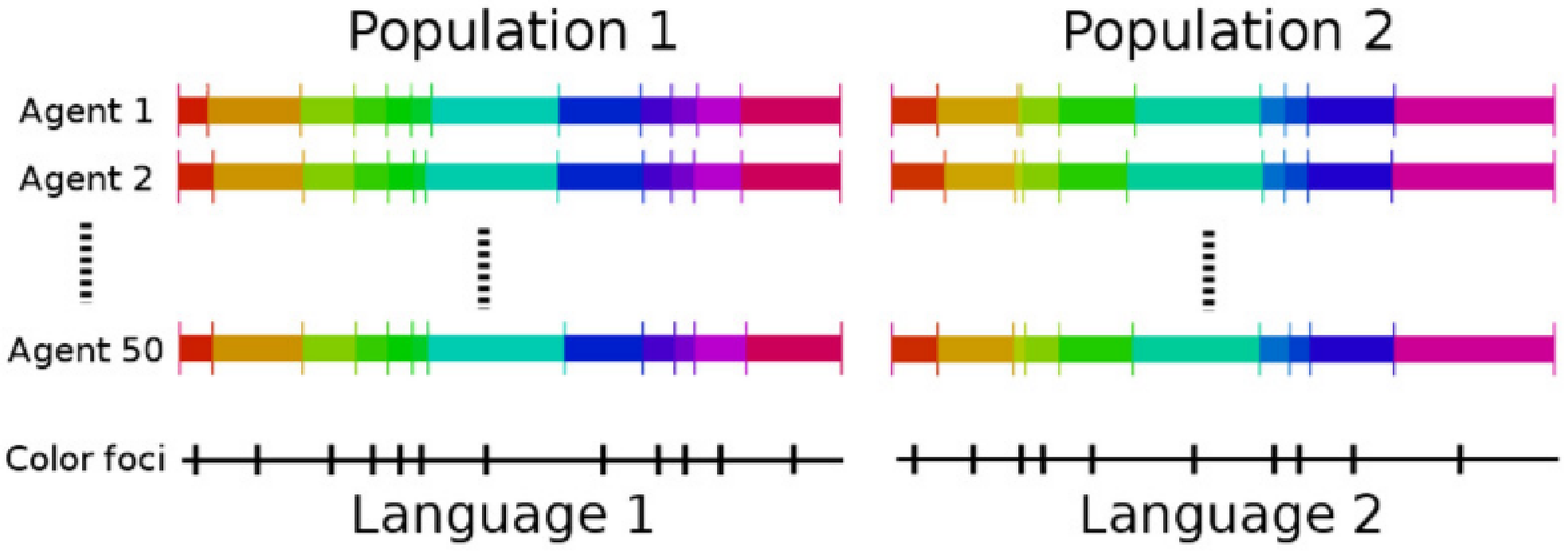}
 \vspace{0.8cm}
 \caption{An example of the results from the
simulations of two different populations with the human JND
($d_{min}(x)$) function. After $10^4$ games, the pattern of categories
and associated color terms are stable throughout the
population. Different individuals in one population have slightly different
category boundaries, but the agreement is almost perfect (larger than
$90\%$). As for each category, a focal color point is defined as the
average of the midpoints of the same category across individuals in the
population. Different populations may develop different final
patterns.}
 %\label{fig:robust_hist}
 \end{center}
 \end{figure}

\begin{figure}
 \begin{center}
 \includegraphics[width=14cm,clip=true]{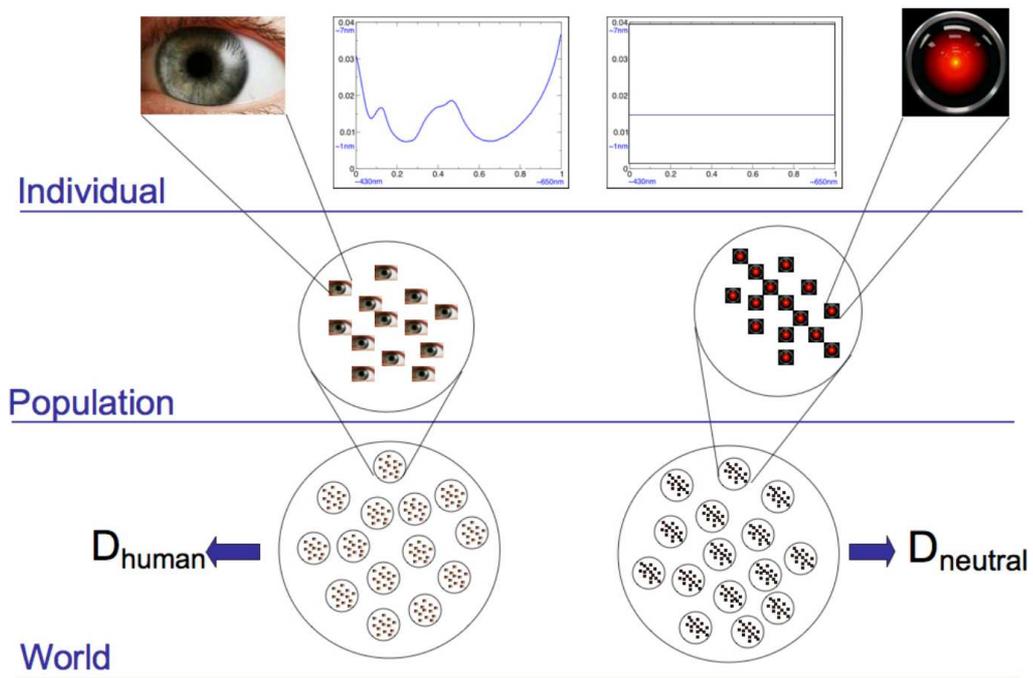}
 \vspace{0.8cm}
 \caption{A sketch of the logical structure of the
Numerical World Color Survey. A value of dispersion $D$ is computed
for each world. A world is an ensemble of populations; each population
achieves a final pattern of color-names shared by its members, and
each individual is endowed with a JND function $d_{min}(x)$. In a
''human world'' (left), all individuals have the human $d_{min}(x)$;
in a ''neutral world'' (right) all individuals have a flat
$d_{min}=0.0143$.}
 %\label{fig:robust_hist}
 \end{center}
 \end{figure}
 
 \begin{figure}
 \begin{center}
 \includegraphics[width=14cm,clip=true]{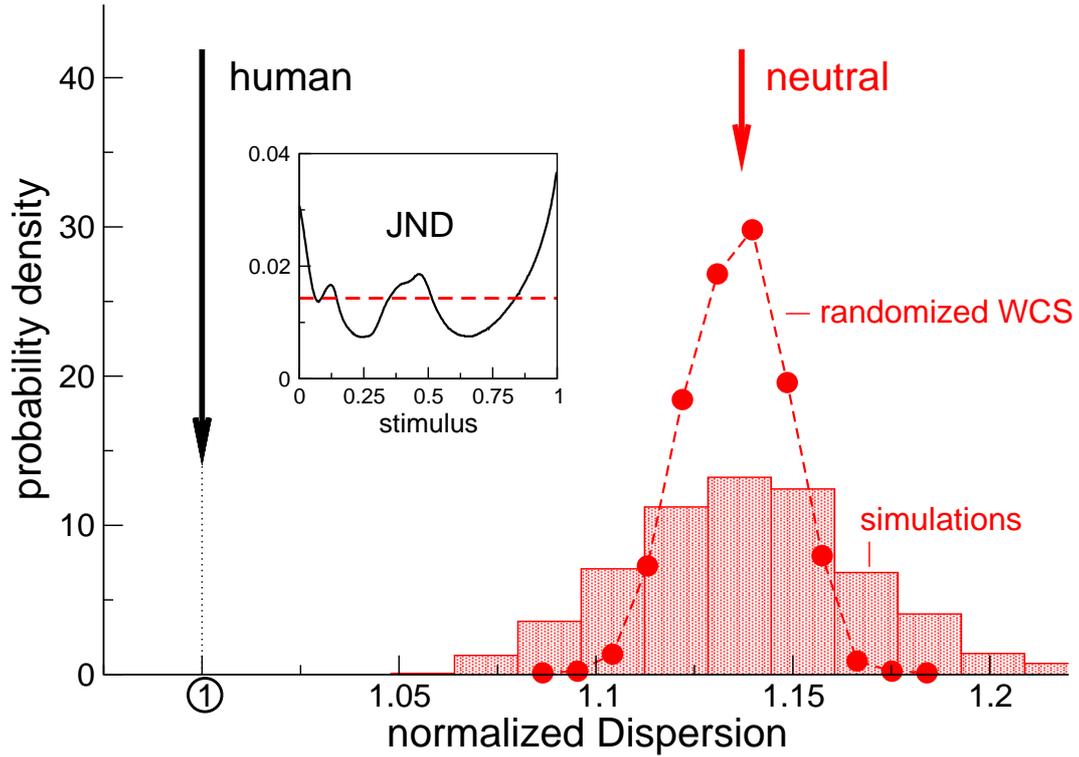}
 \vspace{0.8cm}
 \caption{The dispersion of the ``neutral worlds'',
$D_{neutral}$, (histogram) is significantly higher than that of the
``human worlds'', $D_{human}$, (black arrow), as also observed in the
WCS data (the filled circles extracted from~\cite{kay2003rqc} and the
black arrow). The abscissa is rescaled so that the human $D$ (WCS) and
the average ``human worlds'' $D$ both equal $1$. The histogram has been generated 
from $1500$ neutral worlds, each made of $50$ populations
of $50$ individuals, and $M=2$. The inset figure is the human JND
function (adapted from~\cite{long2006ssn}).}
 %\label{fig:robust_hist}
 \end{center}
 \end{figure}
\end{document}